\documentclass{sig-alternate-2013}
\usepackage[breaklinks]{hyperref}
\usepackage{url}
\usepackage{listings}
\usepackage{color}
\definecolor{gray}{rgb}{0.4,0.4,0.4}
\definecolor{darkblue}{rgb}{0.0,0.0,0.6}
\definecolor{cyan}{rgb}{0.0,0.6,0.6}
\definecolor{red}{rgb}{0.6,0.0,0.0}

\lstdefinelanguage{XML}
{
 morestring=[b]",
 morestring=[s]{>}{<},
 morecomment=[s]{<?}{?>},
 stringstyle=\color{black},
 identifierstyle=\color{darkblue},
 keywordstyle=\color{cyan},
 morekeywords={capability,change,hash,length,type,rel,href,modified},
 emph={rs,link,ln,meta,md,size,mimetype,fixity},
 emphstyle=\color{red}
}

\begin{document}
\title{Extending Sitemaps for ResourceSync}
\numberofauthors{2} 
\author{
% 1st. author
\alignauthor
Martin Klein\\
       \affaddr{Los Alamos National Laboratory}\\
       \affaddr{NM, USA}\\
       \email{mklein@lanl.gov}
% 2nd. author
\alignauthor
Herbert Van de Sompel\\
       \affaddr{Los Alamos National Laboratory}\\
       \affaddr{NM, USA}\\
       \email{herbertv@lanl.gov}
}

\maketitle
\begin{abstract}
The documents used in the ResourceSync synchronization framework are based on the widely adopted document format
defined by the Sitemap protocol.
In order to address requirements of the framework, extensions to the Sitemap format were necessary.
This short paper describes the concerns we had about introducing such extensions, the tests we did to evaluate their 
validity, and aspects of the framework to address them.
\end{abstract}
\section{Introduction}
The ResourceSync collaboration between the National Information Standards Organization (NISO) and the 
Open Archives Initiative (OAI) focuses on designing an approach for the synchronization of web resources.
As resources constantly change (being created, updated, deleted, and moved) 
\cite{cho:estimating,fetterly:large,ntoulas:whats-new}
applications that leverage them would benefit from a standardized synchronization framework
%that is 
aligned with the Web Architecture \cite{webarch}.
The ResourceSync specification \cite{klein:resync_spec} (in beta phase at the time of writing) fulfills the needs of 
different communities and is, amongst others, 
targeted at cultural heritage portals such as Europeana\footnote{\url{http://www.europeana.eu/}}, repositories of scientific 
articles such as arXiv\footnote{\url{http://arxiv.org}}, and linked data applications such as DBpedia 
Live\footnote{\url{http://live.dbpedia.org/}}.
In the framework we refer to a \textbf{Source} as a server that hosts resources subject to synchronization and to a 
\textbf{Destination} as a system that retrieves those resources to synchronize itself with the Source.

Different use cases imply distinct characteristics. From the Source's perspective the resource volume and the resource 
change frequency are most relevant, whereas synchronization latency and accuracy requirements are essential considerations
for Destinations.
The ResourceSync framework therefore offers multiple modular capabilities that a Source can selectively implement to address
specific synchronization needs. 
For the purpose of this paper we discuss only two of the capabilities. An extensive 
discussion about the theoretical background of the framework can be found in Van de Sompel et al. \cite{vdsomp:resync_dlib}
and we refer to the specification document \cite{klein:resync_spec} for a detailed description of all capabilities.

The here described capabilities are the \textbf{Resource List} and the \textbf{Change List}. 
The Resource List, as the name implies, is a list of resources and their descriptions that the Source makes available 
for synchronization. The Resource List presents a snapshot of the Source's resources at one particular point in time
and a Source can publish a Resource List recurrently e.g., once a week or once a month.
The Change List is a list that provides information about changes to the Source's resources. Depending on the 
publication and update frequency of the Change List, this capability can help decrease synchronization latency and 
reduce communication overhead. It is up to the Source to determine the publication frequency as well as the temporal 
interval that is covered by a Change List. It may, for example, describe all resource changes of one day, or one hour, 
or may simply contain a fixed number of changes, regardless of how long it takes to accumulate them.
Both Resource List and Change List serve the following purposes:
\begin{itemize}
\item Synchronization: allow Destinations to obtain current resources; requires the resources' URI.
\item Audit: allow Destinations to verify the accuracy of their synchronized content; requires the resources' 
last modification date and fixity information.
\item Link: allow Sources to express alternate ways for Destinations to retrieve content; requires inclusion of links.
\end{itemize}
One example for the inclusion of such links is a Source providing a pointer to a mirror location. In this case the Source 
prefers Destinations to obtain the resource from that specified location and not from the original URI, with the intention to 
reduce the load on the Source.
Another example is a Source providing a pointer to partial content, meaning only the part of the resource that has actually
changed. A Destination can obtain this information and use it to patch its local copy of the resource. 
\subsection{Considering Sitemaps}
For consistency and to minimize the barrier of adoption, it is desirable to implement all capabilities based on a single
document format instead of using multiple formats within one framework. 
Sitemaps \cite{sitemap_protocol} serve the purpose of advertising a server's resources and support search engines in 
their discovery. In this sense a Sitemap is fairly similar to a Resource List, which motivated us to investigate the use of 
the Sitemap format for the ResourceSync framework.

A Sitemap is an XML document, which must contain three XML elements: the root element \textit{<urlset>},
one \textit{<url>} child element per resource, and exactly one \textit{<loc>} element as a child of each \textit{<url>} element
that conveys the URI of the resource.
The Sitemap schema \cite{sitemap_schema}, allows only the \textit{<url>} element to be a child of the root element and none
of the mandatory elements can have attributes. Listing \ref{ls:sm_simple} shows a simple Sitemap.
\begin{lstlisting}[language=XML,caption=A Sitemap listing one resource, label=ls:sm_simple]
<?xml version="1.0" encoding="UTF-8"?>
<urlset xmlns="http://www.sitemaps.org/schemas/sitemap/0.9">
  <url>
    <loc>http://example.com/res1</loc>
    <lastmod>2013-02-01T13:00:00Z</lastmod>
  </url>
</urlset>
\end{lstlisting}
The \textit{<loc>} element can be utilized in the ResourceSync framework to convey the URI of the resource that is 
subject to synchronization. Its last modification time can be provided with the Sitemap-native optional \textit{<lastmod>} 
child element of the \textit{<url>} element.
\section{ResourceSync Sitemap Extensions}
In order to use Sitemaps for all ResourceSync capabilities, extensions to the Sitemap format are required on two levels:

At the \textit{resource level}, where child elements of an \textit{<url>} element need to be included that provide:
\begin{itemize}
\item Audit: metadata about the resource such as the nature of the change, the resource's size, its
content-based hash value, MIME-Type etc., and 
\item Link: links to related resources such as mirror locations or partial content.
\end{itemize}
At the \textit{root level}, where child elements of the \textit{<urlset>} element need to be included that provide:
\begin{itemize}
\item Audit: the document's last modification time, 
\item Link: links for navigational support for Destinations and to convey information about the Source, and
\item an indication of the capability implemented by a specific Sitemap document since all capability documents have 
the same format.
\end{itemize}
In general, we wanted to avoid Google consuming ResourceSync documents in an unintended way because it does 
not understand the extensions, yet keep them valid from the perspective of the Sitemap XML schema and 
keep the door wide open for when Google understands and eventually consumes them.
These intentions led to the following concerns:\\
\textbf{Concern 1:}
It was unclear how Google would act upon the inclusion of additional child elements to the \textit{<url>} element. 
Since the Sitemap schema allows for external child elements as long as they are properly declared in their namespace, 
we did not anticipate major issues but still had to convince ourselves that our extensions were compliant.\\
\textbf{Concern 2:}
How Google responds to the inclusion of links, as children of the \textit{<url>} element was unclear. 
If Google, for example, indexes both the URI provided in the \textit{<loc>} element and the URI provided in the link to a mirror
location, it would go against the Source's intention to reduce its load. Also, it is meaningless if Google indexes the URI 
of the partial content as it is not helpful without the full resource pointed to in the \textit{<loc>} element.
These would be unintended consequences of the inclusion of links.\\
\textbf{Concern 3:}
Since the schema does not allow for child elements to the \textit{<urlset>} root element other than \textit{<url>}, 
the concern was that Google would reject the ResourceSync capability documents.
The behavior towards included link elements on this level was unclear too.
\section{Testing Sitemap Extensions}
We conducted a series of informal experiments to determine how Google, as a major search engine, responds to ResourceSync
enhanced Sitemaps. We submitted Sitemaps with varying degrees of modification to Google's Webmaster Tool \cite{google_wmt},
analyzed its immediately returned parsing report, and observed the effects of our Sitemaps to their index.
\subsubsection*{Metadata Additions at the Resource Level}
The goal of this experiment was to test the addition of metadata elements to each \textit{<url>} child element of the 
\textit{<urlset>} root element. 
To convey the type of change the resource underwent, we tested the addition of an \textit{rs:change} attribute to the 
\textit{<lastmod>} element.
For additional metadata, we tested new elements in the ResourceSync namespace such as \textit{<rs:size>}, 
\textit{<rs:fixity>}, and \textit{<rs:mimetype>} to convey the resource's size, content-based hash value, and 
MIME-Type, respectively.
Listing \ref{ls:change_list} shows a Sitemap-based Change List that we tested against Google.
\begin{lstlisting}[language=XML,caption=A Change List with added metadata, label=ls:change_list]
<?xml version="1.0" encoding="UTF-8"?>
<urlset xmlns="http://www.sitemaps.org/schemas/sitemap/0.9"
        xmlns:rs="http://www.openarchives.org/rs/terms/">
  <url>
    <loc>http://example.com/res1</loc>
    <lastmod rs:change="updated">
      2013-01-02T13:00:00Z
    </lastmod>
    <rs:size>6230</rs:size>
    <rs:fixity type="md5">
      a2f94c567f9b370c43fb1188f1f46330
    </rs:fixity>
    <rs:mimetype>text/html</rs:mimetype>
  </url>
</urlset>
\end{lstlisting}
All tested child elements from the ResourceSync namespace were tolerated and the \textit{rs:change} attribute, even though in 
violation to the Sitemap schema, was ignored.

However, even though this approach proved feasible, we decided against the addition of multiple child elements 
and in favor of just one additional child element with multiple attributes. 
We named the child element \textit{<rs:md>} and the possible attributes to describe a resource in a Change List are 
\textit{change}, \textit{length}, \textit{hash}, and \textit{type} conveying the same metadata as above.
This approach has two main advantages. First, there is only one added child element that needs to be defined in 
the ResourceSync namespace and secondly, its attributes are defined in the Atom Syndication Format \cite{rfc4287} 
and the Atom Link Extension Internet Draft \cite{snell:atom_link_extension}. Their semantics are inherited in the 
ResourceSync framework.
\subsubsection*{Link Additions at the Resource Level}
%to the \textit{<url>} Block}
%
To provide links to related resources we tested the \textit{<rs:link>} element from the ResourceSync namespace
with the URI being conveyed in its \textit{href} attribute as seen in Listing \ref{ls:change_list_link}.
To provide a mirror location, the link has the relation type \textit{duplicate} (defined in RFC6249 \cite{rfc6249}) and for
partial content (for example JSON patch \cite{json-patch}) a patch-specific relation type. 
%\textit{http://www.openarchives.org/rs/terms/patch}.
%
\begin{lstlisting}[language=XML,caption=A Change List with related resource links, label=ls:change_list_link]
<?xml version="1.0" encoding="UTF-8"?>
<urlset xmlns="http://www.sitemaps.org/schemas/sitemap/0.9"
        xmlns:rs="http://www.openarchives.org/rs/terms/">
  <url>
    <loc rel="nofollow">http://example.com/res1</loc>
    <lastmod>2013-01-02T13:00:00Z</lastmod>
    <rs:link rel="duplicate"
             href="http://mirror.example.com/res1"/>
    <rs:link rel="http://www.openarchives.org/rs/terms/patch"
             href="http://example.com/res1-json-patch"
             type="application/json-patch"/>
  </url>
</urlset>
\end{lstlisting}
Google did not return an error but we did observe unintended consequences (concern 2) with this approach as we 
found both linked resources (\url{http://mirror.example.com/res1} and \url{http://example.com/res1-json-patch}) indexed. 
Our informal tests indicate that Google parses Sitemaps aggressively and indexes URIs it discovers.
For a resource synchronization framework this can be a real detriment because resources in a Resource List or Change List are 
subject to synchronization but they may not be meant for indexing by search engines. 
To address this concern, we tested the \textit{rel="nofollow"} attribute in the \textit{<loc>} child element as well as in the 
\textit{<rs:link>} child elements with the goal of preventing Google from indexing the referenced resource. However, the 
attribute was ignored in either child element. It did not cause any warnings or errors but it also did not prevent Google 
from indexing the resource. 
We were able to improve on this situation by renaming the child element to something different than \textit{<link>} and include 
the URI as its content rather then the value of its \textit{href} attribute. 
However, we adopted the former approach because expressing a link without using the \textit{href} attribute is counterintuitive.

The resulting approach provides no guarantees that Google will not index the URIs provided in links.
Therefore, we additionally introduced an approach that separates discovery of ResourceSync capability documents from discovery
of regular Sitemaps.
%
%
%However, we propose the following approach in the ResourceSync framework:
%A Source should isolate the discovery of the ResourceSync capabilities from a regular Sitemap it may offer. 
We define a \textbf{Capability List} as a document that lists links to all capability documents offered by a Source.
Unlike a Sitemap, which is usually discovered via the \textit{robots.txt} file, the Capability List is discovered via the 
well-known URI \textit{./well-known/resourcesync}, as defined in the ResourceSync specification \cite{klein:resync_spec}.
This distinct discovery is a best effort approach to implement a separation of concerns but there is no guarantee that Google 
does not discover the well-known URI and follow the links to the ResourceSync capability documents.
We would be happy to see search engines such as Google adopting the ResourceSync format but as long as they do not 
understand how to interpret the content of the capability documents, the Source might be better off not to advertise 
them in the robots.txt.

An interesting aspect of the parsing of the Change List shown in Listing \ref{ls:change_list_link} was that Google 
returns a warning that it expects link elements (as well as ``meta'' elements) to be in the XHTML namespace. 
None of the above results changed when using the \textit{<xhtml:link>} child element from the XHTML namespace and 
so, to remain within the ResourceSync namespace, we renamed the link element to \textit{<rs:ln>}.
\subsubsection*{Capability Distinction and Last Modification at the Root Level}
To help Destinations distinguish between capability documents and to convey the document's last 
modification time we tested the 
insertion of the \textit{<rs:md>} child element to the \textit{<urlset>} root element with two attributes.
The attribute identifying the capability document is called \textit{capability} 
(as defined in the ResourceSync specification \cite{klein:resync_spec}) 
and the document's last modification time is conveyed with the attribute \textit{modified}.  
We included the child element into a Resource List (Listing \ref{ls:cap_id}) and submitted the document 
to Google.
\begin{lstlisting}[language=XML,caption={Resource List with capability identifier and last modification date}, label=ls:cap_id]
<?xml version="1.0" encoding="UTF-8"?>
<urlset xmlns="http://www.sitemaps.org/schemas/sitemap/0.9"
        xmlns:rs="http://www.openarchives.org/rs/terms/">
  <rs:md capability="resourcelist"
          modified="2013-02-03T09:00:00Z"/>
  <url>
    <loc>http://example.com/res1</loc>
    <lastmod>2013-02-01T13:00:00Z</lastmod>
  </url>
</urlset>
\end{lstlisting}
Google did not reject the Sitemap, even tough it violated the XML schema. 
It merely returned a warning that the child element is not recognized. 
This supports our suspicion that the Google does not validate submitted Sitemaps against the
schema but rather uses a different logic, which we can only speculate about, to evaluate its correctness.
\subsubsection*{Links at the Root Level}
Two kinds of links at the root level of a ResourceSync document are featured in the framework. 
A navigational link pointing to the Capability List to support Destinations in discovering all offered
capabilities and a link to a document that provides information about the Source.
\begin{lstlisting}[language=XML,caption=Sitemap with navigational and informational links, label=ls:nav_link]
<?xml version="1.0" encoding="UTF-8"?>
<urlset xmlns="http://www.sitemaps.org/schemas/sitemap/0.9"
        xmlns:rs="http://www.openarchives.org/rs/terms/">
  <rs:ln rel="resourcesync"
         href="http://example.com/capabilitylist.xml"/>
  <rs:ln rel="describedby"
         href="http://example.com/info-about-source.xml"/>
  <url>
    <loc>http://example.com/res1</loc>
    <lastmod>2013-02-01T13:00:00Z</lastmod>
  </url>
</urlset>
\end{lstlisting}
We tested this idea and included two \textit{<rs:ln>} child elements from the ResourceSync namespace into the 
submitted Sitemap. The link to the Capability List has the relation type \textit{resourcesync} 
(defined in \cite{klein:resync_spec}) 
and the informational link has the relation type \textit{decribedby} (as defined in 
POWDER \cite{powder}). Listing \ref{ls:nav_link} shows the structure of the Sitemap used for this experiment.
Google did not reject the Sitemap, even though it contains child 
elements of the \textit{<urlset>} element different than \textit{<url>}. It did return a warning though that the child elements 
are not recognized. Unlike in our experiment with links in a \textit{<url>} block, the URIs of these links were not indexed. 
\section{Summary}
%\section{Discussion and Conclusions}
%
%
The purpose of this series of experiments was to test our Sitemap format extensions and to see how Google would
respond to them when submitted to their Webmaster Tool.\\
\textbf{Concern 1}
did not materialize. The Sitemap schema allows for external elements within the \textit{<url>} block 
and hence these extensions are perfectly compliant.\\
\textbf{Concern 2}
did materialize as we saw unintended consequences in terms of indexed URIs that were provided with link
elements. Our tests indicate that Google is rather aggressive in indexing URIs from link elements as they occur in 
\textit{<url>} blocks.
We approach this situation by isolating the discovery of ResourceSync capabilities (via the ResourceSync specific
well-known URI) from regular Sitemaps (via robots.txt). The well-known URI refers to a Capability List containing pointers
to all offered capability documents.\\
\textbf{Concern 3}
did not materialize. Even though the schema did not allow for child elements of the \textit{<urlset>} root
element, Google did not reject our syntax. We suspect that Google does not validate a submitted Sitemaps against the schema 
but rather uses some unknown logic to evaluate the correctness of the Sitemaps. 
In addition, our conversations with Microsoft and Google resulted in their adjustment of the Sitemap schema \cite{sitemap_schema} 
to allow for child elements to the root element. This means that the ResourceSync enhancements to Sitemaps are now fully compliant.
URIs provided in link elements on this level were not subject to be indexed.
Listing \ref{ls:change_list_complete} shows a Change List based on the Sitemap format as adopted in the specification.
%The ResourceSync format is based on Sitemaps but it implies their extension by two elements: \textit{<rs:ln>} and \textit{<rs:md>}. 
%
\begin{lstlisting}[language=XML,caption=A ResourceSync Change List, label=ls:change_list_complete]
<?xml version="1.0" encoding="UTF-8"?>
<urlset xmlns="http://www.sitemaps.org/schemas/sitemap/0.9"
        xmlns:rs="http://www.openarchives.org/rs/terms/">
  <rs:ln rel="resourcesync"
         href="http://example.com/capabilitylist.xml"/>
  <rs:ln rel="describedby"
         href="http://example.com/info-about-source.xml"/>
  <rs:md capability="changelist"
          modified="2013-02-03T09:00:00Z"/>
  <url>
    <loc>http://example.com/res1</loc>
    <lastmod>2013-01-02T13:00:00Z</lastmod>
    <rs:md change="updated"
            length="6230"
            type="text/html"
            hash="md5:a2f94c567f9b370c43fb1188f1f46330"/>
    <rs:ln rel="duplicate"
           href="http://mirror.example.com/res1"/>
    <rs:ln rel="http://www.openarchives.org/rs/terms/patch"
           href="http://example.com/res1-json-patch"
           type="application/json-patch"/>
  </url>
</urlset>
\end{lstlisting}
We also tested the Atom Syndication Format and even introducing a ResourceSync-specific document format as alternatives 
to the Sitemap format. Our reasoning for the decision in favor of the Sitemap format is detailed in our previous work Klein et 
al. \cite{klein:resync_dlib}.
We did not run extensive tests with other search engines. While this is subject to future work, initial tests indicate 
that Microsoft's Bing, for example, is even more liberal in accepting our Sitemap extensions.
\section{Acknowledgments}
The ResourceSync specification is the collaborative work of NISO and OAI. 
Funding is provided by the Alfred P. Sloan Foundation and UK participation is supported by Jisc.
%
%\bibliographystyle{abbrv}
%\bibliography{mklein_jcdl2013}
%

%
% That's all folks!
\end{document}